\documentclass[3p,times,procedia]{elsarticle}
\usepackage{nupha_ecrc}
\usepackage[utf8x]{inputenc}

\usepackage{hyperref}
\volume{00}

\firstpage{1}

\journalname{Nuclear Physics A}

\runauth{H. Mäntysaari et al.}

\jid{nupha}

\jnltitlelogo{Nuclear Physics A}

\usepackage{amssymb}
\usepackage{amsmath}
\newcommand{\rt}{{\mathbf{r}}}
\newcommand{\xt}{{\mathbf{x}}}
\newcommand{\nc}{N_\mathrm{c}}
\newcommand{\bt}{{\mathbf{b}}}
\newcommand{\bti}{{\mathbf{b}_{i}}}
\newcommand{\gev}{\ \textrm{GeV}}
\newcommand{\der}{\mathrm{d}}
\newcommand{\xpom}{{x_\mathbb{P}}}
\newcommand{\A}{\mathcal{A}}

\usepackage[figuresright]{rotating}

\begin{document}

\begin{frontmatter}



%% Instructions from Editor: Please use the following \dochead only in the preprint version (e-print arXiv etc.); 
%% use empty \dochead{} when submitting to Nuclear Physics A!
%\dochead{XXVIth International Conference on Ultrarelativistic Nucleus-Nucleus Collisions\\ (Quark Matter 2017)}
\dochead{}
%% e.g. \dochead{17th International Conference on Dynamical Processes in Excited States of Solids}

\title{Proton structure fluctuations: constraints from HERA and applications to $p+A$ collisions}

\author{Heikki Mäntysaari}
\author{Björn Schenke}
\author{Chun Shen}
\author{Prithwish Tribedy}

\address{Physics Department, Brookhaven National Laboratory, Upton, NY 11973, USA}

\begin{abstract}
We constrain proton structure fluctuations at small-$x$ by comparing with the HERA diffractive vector meson production data, and find that large geometric fluctuations in the proton wave function are needed. Hydrodynamical simulations of proton-nucleus collisions including the fluctuating proton structure are found to be compatible with the
momentum anisotropies measured at the LHC.
% LHC flow harmonic measurements.
%When the fluctuating protons are used to simulate proton-nucleus collisions, we obtain harmonic flow coefficients comparable with the LHC data.
\end{abstract}


\end{frontmatter}

% \linenumbers

\section{Introduction}
\label{intro}

One of the striking findings of the proton-nucleus collisions at the LHC has been the presence of strong collective phenomena in high multiplicity events. These signals, 
%like elliptic flow and long range rapidity correlations,
particularly long range rapidity correlations with distinct anisotropies in the azimuthal direction,
 are very similar to those that have been traditionally 
 interpreted as evidence for the creation of an almost perfect fluid
% associated as signatures of the Quark Gluon Plasma
  in ultrarelativistic heavy ion collisions (for a review of recent collectivity measurements, see Ref.~\cite{Dusling:2015gta}). This has raised a natural question whether the high-multiplicity proton-nucleus, and even proton-proton collisions, can be described using the same relativistic hydrodynamic framework that has been successful in describing a vast variety of LHC and RHIC heavy ion data (see e.g. Ref~\cite{Schenke:2012wb}).

In hydrodynamical simulations the initial state geometric anisotropies are converted into momentum space correlations and are observed e.g. as an elliptic ($v_2$) and triangular flow ($v_3$). Thus, in order to model proton-nucleus collisions using relativistic hydrodynamics, a detailed knowledge of the initial state geometry is crucial. In particular, the event-by-event fluctuations of the (on average round) proton are needed to model the initial state. Without large eccentricities in the initial state it is difficult to explain the large flow harmonics observed at the LHC~\cite{Schenke:2014zha}. 

One possibility to constrain the proton structure fluctuations is given by diffractive deep inelastic scattering measured at the HERA electron-proton collider. Unlike other DIS observables, diffractive scattering processes are sensitive to the transverse geometry of the target. The actual process we have in mind here is exclusive vector meson production (in this work $J/\Psi$). These processes can be divided into two categories. In \emph{coherent diffraction}, where the target hadron remains in the same quantum state, the transverse momentum spectra of the produced vector mesons are directly related to the average density profile of the target~\cite{Miettinen:1978jb,Kowalski:2006hc}. On the other hand, in \emph{incoherent diffraction}, where the target breaks up but there is still no net color charge exchanged between the vector meson and the target, the cross section is proportional to the \emph{amount of fluctuations} in the target wave function~\cite{Miettinen:1978jb,Frankfurt:2008vi,Lappi:2010dd}. 

In this work, we constrain the proton shape fluctuations using  HERA diffractive $J/\Psi$ production data, and use the obtained fluctuating protons as an input for hydrodynamical simulations of proton-nucleus collisions at the LHC energy $\sqrt{s_{NN}}=5.02\,\mathrm{TeV}$. The diffractive calculations are published in Refs.~\cite{Mantysaari:2016ykx,Mantysaari:2016jaz} and the hydrodynamical simulations in Ref.~\cite{Mantysaari:2017cni}. 

\section{Theoretical framework: IP-Glasma}
\label{sec:fluctuations}
Coherent vector meson production data from HERA has been described well by  dipole model calculations (see e.g.~\cite{Rezaeian:2012ji}), assuming a Gaussian density profile for the proton, $T_p(\bt) = \frac{1}{2\pi B} e^{-\bt^2/(2B)}$,
with $B=4 \gev^{-2}$. We include event-by-event fluctuations by assuming a constituent quark inspired picture, where the small-$x$ gluons are located around the three valence quarks. We sample the locations of these \emph{hot spots} in the transverse plane from a Gaussian distribution that has a width $B_{qc}$. The density profile of each of these hot spots is then assumed to be Gaussian with different width $B_q$. This corresponds to the replacement
%\begin{equation}
$T_p(\bt) \to \frac{1}{3} \
\sum_{i=1}^3 T_q(\bt-\bti)$ with %,  \quad \text{with }
$ T_q(\bt) = \frac{1}{2\pi B_q} e^{-\bt^2/(2B_q)}$.
%\end{equation}
This replacement is performed in the impact parameter dependent saturation model (IPsat) fitted to inclusive HERA data~\cite{Rezaeian:2012ji}. In addition, we include saturation scale ($Q_s$) fluctuations as described in Ref.~\cite{Mantysaari:2016jaz}.
 From the IPsat model, we can then calculate the saturation scale $Q_s$ at each point in the transverse plane, and relate it to the color charge density.
  After solving the Yang-Mills equations we obtain the Wilons lines at each point in the transverse plane. From the Wilson lines, it is possible to evaluate the dipole-proton amplitude which is a necessary ingredient to calculations of diffractive cross sections as discussed below. The Wilson lines also include all the information about the initial state when we use hydrodynamical simulations to describe proton-nucleus collisions. For more details, the reader is referred to Refs.~\cite{Mantysaari:2016ykx,Mantysaari:2016jaz,Mantysaari:2017cni} and references therein.
 
 \begin{figure}[tb]
    \centering
    \begin{minipage}{.48\textwidth}
        \centering
        \includegraphics[width=0.8\textwidth]{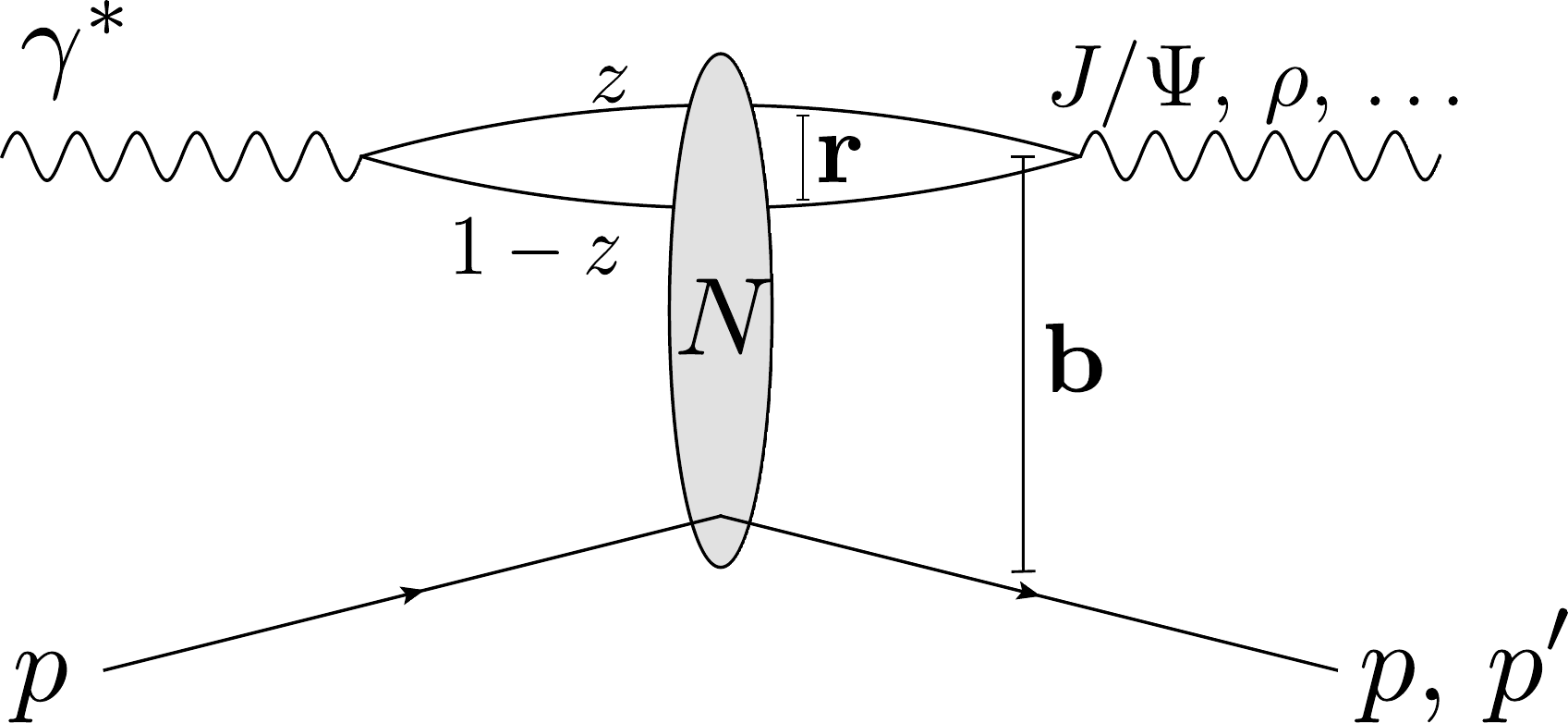} 
				\caption{Diffractive vector meson production in dipole picture.}
		\label{fig:photonproton}
    \end{minipage} \quad
    \begin{minipage}{0.48\textwidth}
        \centering
      \includegraphics[width=\textwidth]{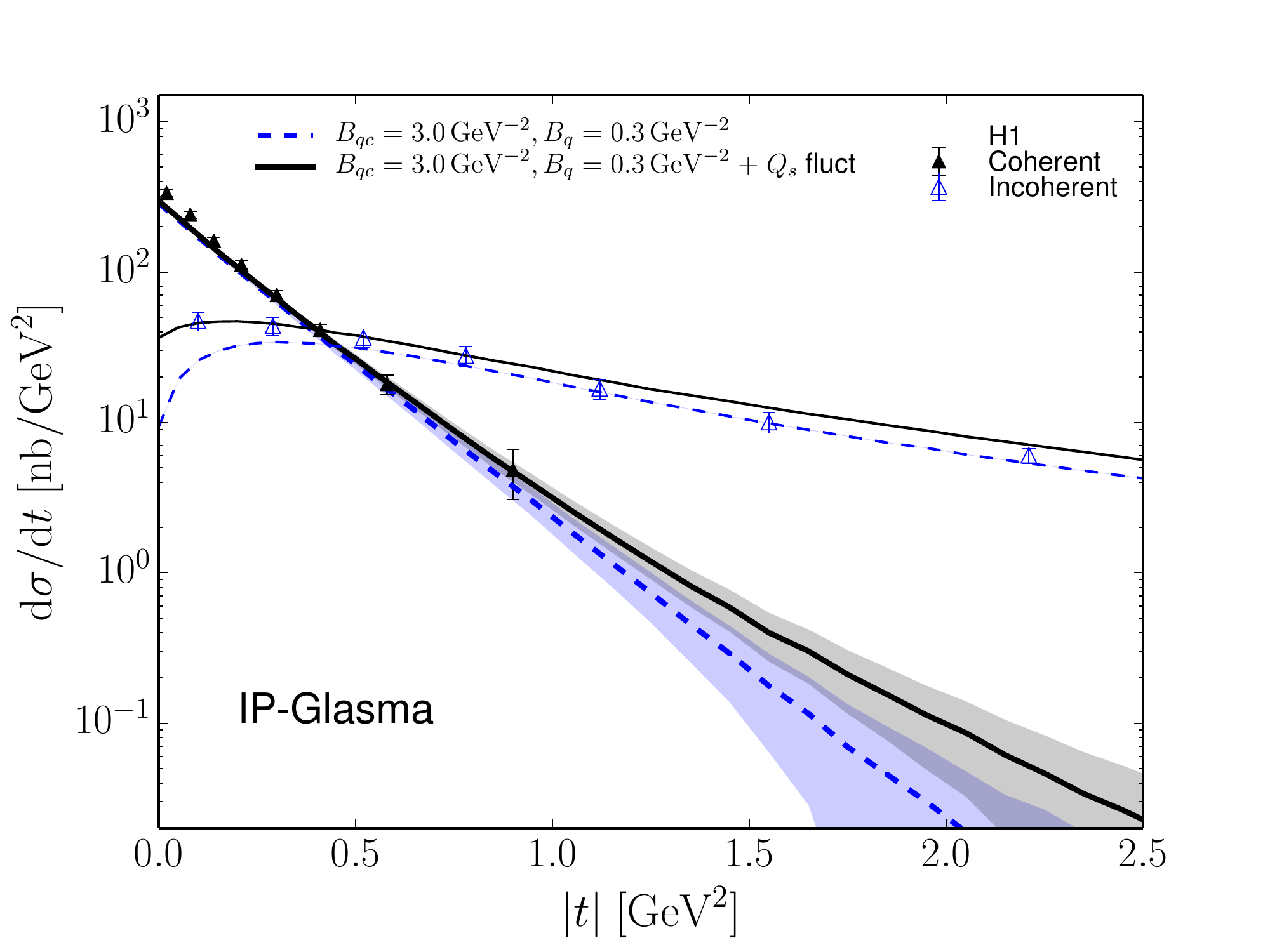} 
				\caption{Coherent and incoherent diffractive $J/\Psi$ production compared with the H1 data~\cite{Alexa:2013xxa}.}
		\label{fig:diffractive_spectra}
    \end{minipage}
\end{figure}

%\section{Diffractive vector meson production}

The scattering amplitude for  diffractive vector meson production 
can be written as~\cite{Kowalski:2006hc}
\begin{equation}
\label{eq:diff_amp}
 \A^{\gamma^* p \to V p}_{T,L}(\xpom,Q^2, \boldsymbol{\Delta}) = 2i\int \der^2 \rt \int \der^2 \bt \int \frac{\der z}{4\pi}  (\Psi^*\Psi_V)_{T,L}(Q^2, \rt,z)  e^{-i[\bt - (1-z)\rt]\cdot \boldsymbol{\Delta}}  N(\bt,\rt,\xpom).
\end{equation}
Here $Q^2$ is the virtuality of the photon, $\bt$ impact parameter, $\rt$ size of the quark-antiquark dipole and $z$ longitudinal momenum fraction of the photon carried by the quark.
% and $\Psi^*\Psi$ describes the overlap between the virtual photon and the vector meson wave functions. 
The transverse momentum of the vector meson $\boldsymbol{\Delta} \approx \sqrt{|t|}$ is the Fourier conjugate to $\bt - (1-z)\rt$. This scattering probes the gluon distribution of the target at longitudinal momentum fraction $\xpom$. The dipole-proton amplitude $N$ is obtained as a trace of Wilson lines at positions $\bt + \rt/2$ and $\bt - \rt/2$.
The interpretation of Eq.~\ref{eq:diff_amp} is straightforward: first, the incoming virtual photon splits to a quark-antiquark dipole. This splitting is described by the virtual photon wave function $\Psi$. The dipole then scatters elastically off the proton, which is given in terms of the dipole scattering amplitude $N$. Finally, the quark-antiquark pair forms a vector meson according to the vector meson wave function $\Psi_V$. This is illustrated in Fig.~\ref{fig:photonproton}.

The coherent diffractive cross section is obtained by averaging the scattering amplitude over different configurations of the proton, and then taking the square:
%\begin{equation}
%\label{eq:coherent}
$\frac{\der \sigma^{\gamma^* p \to V p}}{\der t} = \frac{1}{16\pi} \left| \langle \A^{\gamma^* p \to V p}(\xpom,Q^2,\boldsymbol{\Delta}) \rangle \right|^2.$
%\end{equation}
%where $\A^{\gamma^* p \to V p}(\xpom,Q^2,\boldsymbol{\Delta})$ is the scattering amplitude.
The incoherent cross section can be written as a variance~\cite{Miettinen:1978jb} (see also Refs.~\cite{Frankfurt:2008vi,Lappi:2010dd}): \\
%\begin{equation}\label{eq:incoherent}
$\frac{\der \sigma^{\gamma^* p \to V p^*}}{\der t}  = \frac{1}{16\pi} \left( \left\langle \left| \A^{\gamma^* p \to V p}(\xpom,Q^2,\boldsymbol{\Delta})  \right|^2 \right\rangle - \left| \langle  \A^{\gamma^* p \to V p}(\xpom,Q^2,\boldsymbol{\Delta}) \rangle \right|^2 \right)\,.$
%\end{equation}  

%\section{Hydrodynamical description of the proton-nucleus collisions}
In the hydrodynamical simulations performed using {\sc Music}~\cite{Schenke:2010nt,Schenke:2010rr} we use the same second order transport parameters as in Ref.~\cite{Ryu:2015vwa} and effective constant shear viscosity $\eta/s=0.2$. We also include bulk viscosity and use a temperature dependent $\zeta/s$ as in~\cite{Ryu:2015vwa}, which is essential in small systems to obtain average transverse momenta compatible with the LHC data. Before $\tau_0=0.2\,\mathrm{fm}/c$, when we start hydrodynamical evolution, the system is evolved by solving the classical Yang-Mills equations.  The full energy momentum tensor $T^{\mu\nu}$ at $\tau_0$ is matched to the hydrodynamical phase.
% including the initial shear stress tensor and bulk viscous pressure. 
The hydrodynamic equations are evolved with a Lattice Equation of State s95p-v1~\cite{Huovinen:2009yb}. The switching temperature is set to $T_\text{switch}=155$ MeV, after which particles are sampled using the Cooper-Frye prescription and are allowed to propagate in the hadronic cascade model UrQMD~\cite{Bleicher:1999xi}. For more details, see Ref.~\cite{Mantysaari:2017cni}.

\section{Results}

\begin{figure}[tb]
    \centering
    \begin{minipage}{.48\textwidth}
        \centering
        \includegraphics[width=0.8\textwidth]{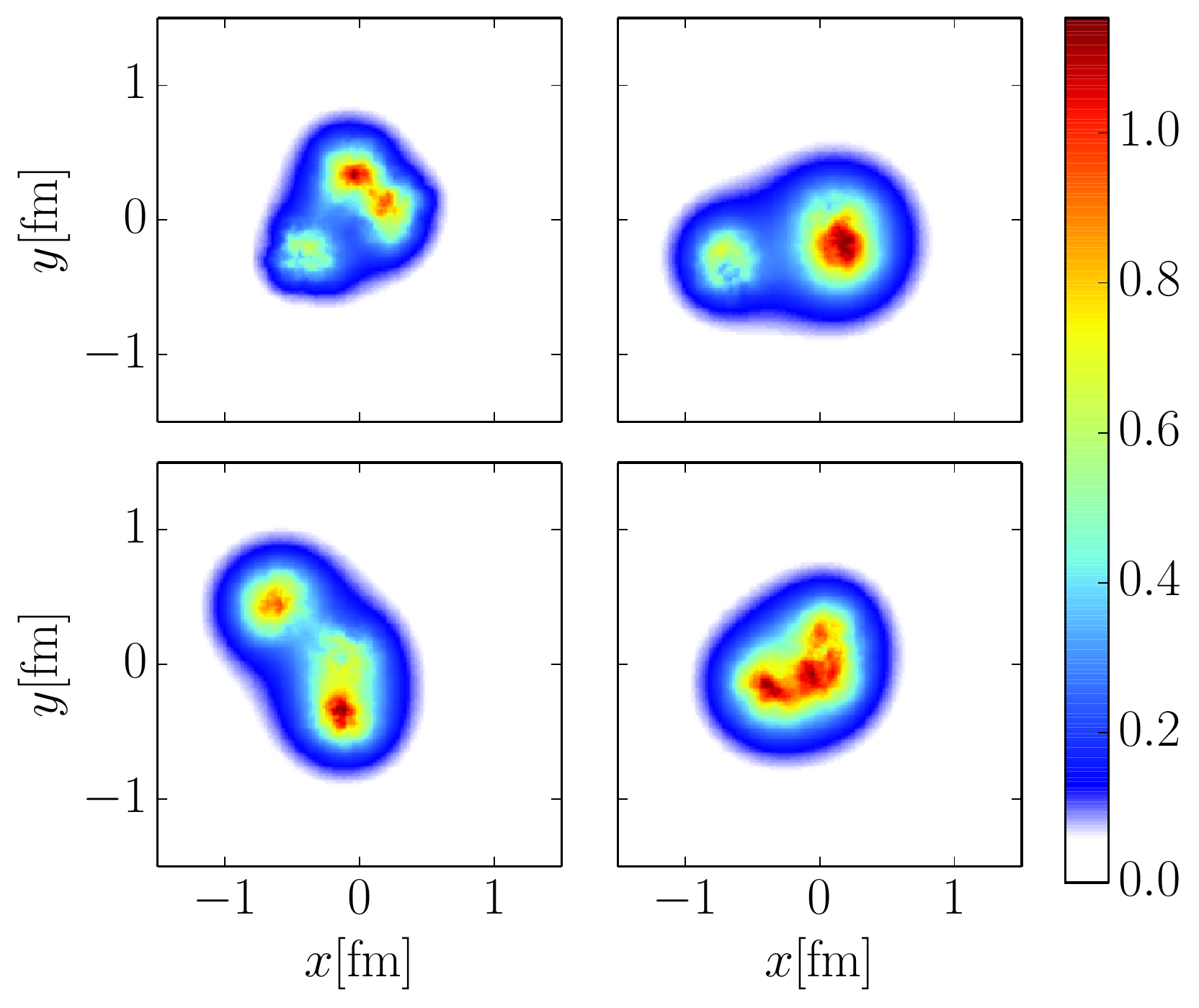} 
				\caption{Example of the proton density profiles at $x \approx 10^{-3}$. The quantity shown is $1 - \mathrm{Re}\, \mathrm{Tr} V(\xt)/\nc$.}
		\label{fig:traces}
    \end{minipage} \quad
    \begin{minipage}{0.48\textwidth}
        \centering
      \includegraphics[width=\textwidth]{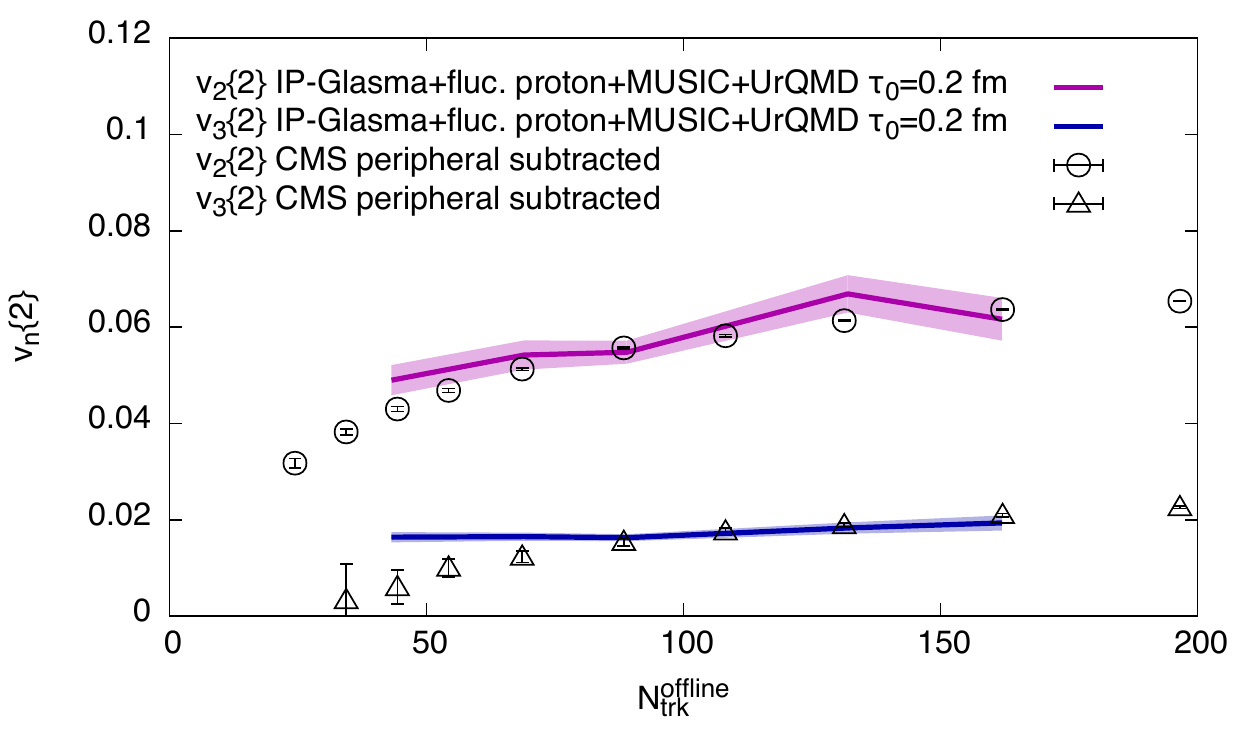} 
				\caption{Elliptic ($v_2$) and triangular ($v_3$) as a function of multiplicity compared to the CMS data~\cite{Chatrchyan:2013nka}. }
		\label{fig:vn}
    \end{minipage}
\end{figure}

First we constrain the amount of fluctuations by comparing with the H1 diffractive $J/\Psi$ photoproduction data~\cite{Alexa:2013xxa}. The free parameters that describe the proton fluctuations, $B_q$ and $B_{qc}$ (see Sec.~\ref{sec:fluctuations}), are obtained by requiring a good simultaneous agreement with both coherent and incoherent diffractive cross sections. As these processes are separately sensitive to the average density profile and to the amount of fluctuations, this allows us to constrain both $B_q = 0.3\gev^{-2}$ and $B_{qc}=3.0\gev^{-2}$. This corresponds to having  large event-by-event fluctuations in the proton wave function. This is illustrated in Fig.~\ref{fig:traces}, where we show $1-\mathrm{Re}\, \mathrm{Tr} V(\xt)/\nc$ which can be seen to be roughly proportional to the density of the proton.

Next, we take the fluctuating proton structure constrained previously and calculate elliptic and triangular flow in proton-lead collisions at $\sqrt{s_{NN}}=5.02$ TeV. The results are compared with the CMS data~\cite{Chatrchyan:2013nka} in Fig.~\ref{fig:vn}. In previous calculations within the same framework but without geometric fluctuations for the proton, the harmonic flow coefficients were largely underestimated~\cite{Schenke:2014zha}. This is in contrast to the results shown in Fig.~\ref{fig:vn}, where a good agreement with the CMS data is obtained at large multiplicities. We also show in Ref.~\cite{Mantysaari:2017cni} that the HBT radii and mean transverse momenta are compatible with the measurements.

\section{Conclusions}
We have constrained the geometric fluctuations of the proton by comparing with the HERA coherent and incoherent diffractive vector meson production data. We find that in order to describe these measurements, large geometric fluctuations of the proton wave function are needed. Simulating proton-nucleus collisions using the fluctuating protons constrained by  HERA data, we find a good description of the flow harmonics in high multiplicity events. In Ref.~\cite{Mantysaari:2017dwh} the necessity of having large geometric fluctuations was also found to improve the description of the exclusive $J/\Psi$ production data in ultraperipheral heavy ion collisions at the LHC.

\section*{Acknowledgments}
This work was supported under DOE Contract No. DE-SC0012704 and used resources of the National Energy Research Scientific Computing Center, supported by the Office of Science of the U.S. Department of Energy under Contract No. DE-AC02-05CH11231. BPS acknowledges a DOE Office of Science Early Career Award.



%% References


\bibliographystyle{elsarticle-num}
\bibliography{../../../refs.bib}






% \end{thebibliography}

\end{document}